\documentclass[11pt, a4paper]{article}
\usepackage[utf8]{inputenc} % Any characters can be typed directly from the keyboard, eg éçñ
\usepackage{textcomp} % provide lots of new symbols
\usepackage{flafter}  % Don't place floats before their definition

\usepackage{amsmath,amssymb}  % Better maths support & more symbols
\usepackage{bm}  % Define \bm{} to use bold math fonts
\usepackage{authblk}
\usepackage{memhfixc}  % remove conflict between the memoir class & hyperref
\usepackage{listings}
\numberwithin{equation}{section}
\begin{document}
\title{Coarse-Grained Stochastic Particle-based Reaction-Diffusion Simulation Algorithm}
\author{Thorsten Pr\"ustel} 
\author{Martin Meier-Schellersheim} 
\affil{Laboratory of Systems Biology\\National Institute of Allergy and Infectious Diseases\\National Institutes of Health}
\maketitle
\let\oldthefootnote\thefootnote 
\renewcommand{\thefootnote}{\fnsymbol{footnote}} 
\footnotetext[1]{Email: prustelt@niaid.nih.gov, mms@niaid.nih.gov} 
\let\thefootnote\oldthefootnote 
\abstract
{
In recent years, several particle-based stochastic simulation algorithms (PSSA) have been developed to study the spatially resolved dynamics of biochemical networks at a molecular scale. A challenge all these approaches have to address is to allow for simulations at cell-biologically relevant timescales without neither neglecting important spatial and biochemical properties of the simulated system nor introducing ad-hoc assumptions not based on physical principles. Here we describe a PSSA that permits large time steps while still retaining a high degree of accuracy. The approach addresses the typical disadvantage of Brownian dynamics, namely the need to use small time steps to resolve bimolecular encounters accurately, by estimating the number of otherwise unnoticed encounters with the help of the Green's functions of the diffusion equation incorporating molecular interactions. This method has previously been proposed for purely absorbing boundary conditions and irreversible bimolecular reactions. Building on those ideas, we developed a general-purpose PSSA that is applicable to a broad class of reaction-diffusion problems by incorporating reflective and radiation boundary conditions and reversible reactions. We furthermore discuss how reaction-diffusion systems on 2D membranes can be described and derive small time expansions of the Green's functions that substantially speed up key calculations, particularly in the problematic case of molecules in close proximity.  Finally, we point out the formal relationship between our and exact algorithms. The proposed algorithm may serve as an easily implementable and flexible, computationally efficient, coarse-grained description of reaction-diffusion systems in 2D and 3D that nevertheless provides a stochastic, detailed representation at the level of individual particle trajectories in space and time.
}

\section{Introduction}
Stochastic fluctuations are inherent to any cellular biochemical system due to its molecular constituents. Under certain conditions, which often include the existence of sufficiently separated timescales between different levels of organization of cellular biochemistry, the fluctuations can be averaged out and one arrives at an effective theory that treats the system at a coarser level of description at which fluctuations are negligible. The relevant level of description is defined by the experimental context. However, if the effect of fluctuations can propagate through the different scales of the system, or if the experimental context changes, a theoretical treatment that explicitly takes into account fluctuations becomes necessary. The chemical master equation (CME) \cite{kampen2007stochastic}, which abandons the notion of molecule concentrations and incorporates stochastic fluctuations, provides the appropriate theoretical framework. Unfortunately, it is difficult if not impossible to solve the CME for all but the simplest reaction networks. The stochastic simulation algorithm (SSA) or "Gillespie algorithm"  \cite{gillespie1976general,gillespie1977exact} avoids the need of finding a solution by sampling numerical realizations of the network components' time evolution according to the CME .  However, intracellular biochemical networks operate in an environment that exhibits a spatially intricate, highly heterogeneous organization with aspects such as compartmentalization and scaffolding playing an important role.  Thus, simulation tools are needed that are capable of abandoning the requirement of a well-stirred and homogeneous environment. The SSA incorporates fluctuations but still assumes that the system is well-stirred and spatially homogenous. These shortcomings can be overcome by the so-called spatial Gillespie algorithms which partition the reaction volume in compartments small enough that within them the assumption of a well-stirred and homogeneous system are justified again \cite{ander2004smartcell, Hattne:2005p86, rodriguez2006spatial}. However, this approach relies on the existence of a length and time scale on which the system is again homogeneous. Such a scale may not always exist for a particular intracellular reaction-diffusion network.
Particle-based methods are capable of taking into account fluctuations and spatial aspects without introducing ad-hoc spatial discretizations. Such methods treat biochemical networks as composed of elementary unimolecular ($A\rightarrow products$) and bimolecular ($A+B\rightarrow products$) reactions. 
A bimolecular $A+B\rightarrow C$ reaction may be depicted as a two-step process \cite{shoup1982role, Rice:1985}:
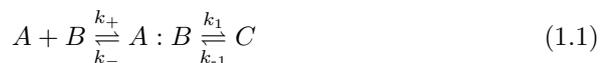
\begin{equation}
A+B\overset{k_{+}}{\underset{k_{-}}{\rightleftharpoons}} A:B \overset{k_{\text{1}}}{\underset{k_{\text{-1}}}{\rightleftharpoons}} C
\end{equation}
According to this kinetic scheme, a prerequisite for the occurrence of a bimolecular reaction is that the two molecules encounter each other through diffusive (Brownian) motion and form an encounter complex $A:B$. The actual chemical reaction is characterized by the rate constant $k_{\text{1}}$. Alternatively, the encounter complex may decay and the molecules escape from each other with the diffusional backward rate $k_{-}$. Finally, the reaction product $C$ can dissociate to the encounter complex with the dissociation constant $k_{\text{-1}}$. 
If the time scale associated with diffusion is comparable or larger than the time scale associated with the intrinsic chemical reaction, the bimolecular reactions are referred to as \emph{diffusion-influenced} or \emph{diffusion-limited} \cite{Rice:1985}. In these cases the Brownian motion of the individual molecules becomes an important element of the theoretical description. 
Conceptually, there are two different approaches to describing Brownian dynamics (BD).
The first approach is based on stochastic differential equations, often referred to as Langevin equations providing a time stepping procedure to generate particular stochastic realizations of a particle's space time trajectory.
The second approach, in contrast, uses the associated Fokker-Planck (FP) equation \cite{risken1996fokker} that governs the deterministic time evolution of the conditional probability density function (pdf) $\rho(\mathbf{r}, t\vert \mathbf{r}_{0}, t_{0})$. The major advantages of the Langevin description are its versatility and the ease of implementing it as a stochastic simulation algorithm.
However, the Langevin method suffers from a severe drawback: Small time steps are in general necessary to resolve the encounter events with sufficient accuracy \cite{BarSchieber:2002BS}. The need for tiny steps becomes especially painful in dilute systems. In these systems, the vast majority of time updates perform purely diffusional steps that do not involve reactions. In principle, this problem can be addressed by the second approach based on the FP equation . The Green's functions of the FP equation contain all relevant information about the time evolution of the distribution of the particles in the system and the FP equation approach has been widely used in the analysis of diffusion-influenced reactions, cp., for instance, \cite{Agmon:1984p31, Agmon:1990p10}. Unfortunately, for many-particle systems analytical solutions are not available. In addition, for more complicated systems it is not feasible either to generate a solution of the Fokker-Planck equation by numerical methods for treating partial differential equations. In contrast, in such systems the associated stochastic differential equation can still be integrated numerically. Indeed, following the seminal work by Ermak and McCammon \cite{Ermak:1978p40} many simulation algorithms based on BD have been employed in a variety of fields, including the study of protein-protein association reactions \cite{Gabdoulline:2002p49}, dynamics of polymeric fluids \cite{ottinger1996stochastic} and more recently, the simulation of networks of interacting biomolecules \cite{Andrews:2004p84,schutter2001computational,morelli2008reaction}.\\
In recent years Green's function based methods have been developed that avoid the problems caused by many-particle systems by determining the time step for every simulation cycle based on the requirement that within $\Delta t$ at most two particles may encounter each other, thereby excluding N-body interactions \cite{vanZon:2005p340, vanZon:2005p401}. Factorizing a many-particle system into several independent 1-body and 2-body systems that can be propagated according to the Green's functions of the 1-body and 2-body diffusion equation permits performing large simulation time steps when the molecular concentrations are small. 
In a similar spirit, the first-passage kinetic Monte Carlo \cite{Opplestrup:2006p235, Oppelstrup:2009p144} reduces the many-particle system to a set of 1-body and 2-body problems by partitioning the reaction volume into protective domains which contain at most two particles. Within the domains the particles can be propagated by first-passage and no-passage Green's functions.
Both methods are more efficient for dilute system than conventional BD simulations. Their event-driven nature ensures that every encounter is taken into account even when using large time steps. While this adds to the accuracy of the algorithms, for higher concentrations and particles close to boundaries the time step can become very small, rendering the methods inefficient.\\
Here, we propose a stochastic simulation framework for reaction-diffusion processes that combines elements of both, the Langevin and Fokker-Planck method.

Briefly, the particles' displacements are sampled according to the free-space (overdamped limit) of the Langevin equation
\begin{equation}\label{oLangevin}
\frac{d \mathbf{r}}{dt} = \frac{D}{k_{B}T}\left(\mathbf{F} + \mathbf{F}_{\text{Brown}}(t)\right).
\end{equation}
where $D, k_{B}, T$ are the diffusion constant, the Boltzmann constant and the absolute temperature, respectively. The stochastic force contribution is characterized by
\begin{equation}
 \langle \mathbf{F}_{\text{Brown}}\rangle = 0 \quad \text{and} \quad \langle F_{\text{Brown},i}(t_{0})F_{\text{Brown},j}(t)\rangle = 2\eta^{2}D\delta_{ij}\delta(t_{0} - t),
\end{equation}
where $\eta = \tfrac{k_{B} T}{D}$ and the deterministic force contribution is described by $\mathbf{F}$.
 To correct for underestimating the number of encounters we employ the 3D and 2D analytical representations of the fundamental solutions of associated Smoluchowski equation with absorbing, reflecting and radiation boundary conditions (bcs). These Green's functions can be used to compute the probability that particles that do not overlap after a time step nevertheless reacted at some intermediate time during the time step.
We emphasize that the general idea to use Green's functions of Fokker-Planck equations to enhance BD simulations is far from new, cp. \cite{Lamm:1983p12, Lamm:1984p4, North:1984, NorthCurv:1986, green1988simulation, pimblott1992stochastic, Edelstein:1993, ottinger1996stochastic}.
Indeed, the method has been proposed for the case of purely absorbing boundary conditions and irreversible reactions already in \cite{BarSchieber:2002BS} . Here, we build on this approach to develop a general purpose simulation framework applicable to a broad class of reaction-diffusion systems by extending it to include reflective and radiation boundary conditions and reversible reactions. Furthermore, we describe how to treat reaction-diffusion systems in 2D. In this case, the radial Green's functions cannot be expressed by elementary functions, in contrast to their 3D counterparts. As a consequence, their numerical approximation is more costly, an issue that becomes worse for smaller time steps. To address this problem, we derive small time expansions for the key expressions. These small time expansions permit to circumvent a numerical integration and should prove useful for any simulation algorithm employing 2D Green's functions.
Finally, we establish a connection between the presented algorithm and the exact first-passage time algorithm to elucidate in what sense it can be understood as coarse-grained version. In this context an anology to the Gillespie formalism is helpful: While the original Gillespie algorithm is event-driven and takes every reaction event into account, resulting in possibly tiny time steps, and therefore rendering it too slow for networks with high molecular abundances, the tau-leaping method \cite{gillespie2001approximate} works with a larger (constant) time step $\Delta t$, but it has to provide a procedure to estimate the number of events that occur during $\Delta t$.

Since the time step that can be taken with our approach depends less sensitively on the concentrations of the chemical species, the method is also applicable to systems for which other stochastic methods, for instance GFRD  \cite{vanZon:2005p340, vanZon:2005p401}, would fail. Compared to those approaches, the time step can be taken much larger while maintaining a high degree of accuracy \cite{BarSchieber:2002BS}. 
Furthermore, chemical reactions near a reflecting boundary (such as a membrane) pose no particular challenge, again in contrast to the mentioned event-driven methods.  Due to its well-defined relationship with event-driven simulation algorithms, cp. section \ref{appendix}, our approach may serve as a complement for situations in which those algorithms are less efficient, for instance when high local particle densities would otherwise require using BD simulation steps \cite{Takahashi:2010p139}. 

A note on terminology: In the following we will always consider the overdamped limit of the Langevin equation, i.e. only position variables, but no velocity variables are taken into account. The corresponding FP equation is referred to as Smoluchowski equation. Furthermore, we will also neglect deterministic force contributions. In this case the Smoluchowski equation takes the form of Einstein's diffusion equation \cite{einstein1956investigations} and in the following we will use the terms diffusion, Smoluchowski and Fokker Planck equation interchangeably. We emphasize that this does not mean a limitation of the approach. Rather, as pointed out by  \cite{BarSchieber:2002BS}, treating the stochastic and deterministic force contributions separately, offers the advantage that it is sufficent to deal only with the Green's functions of the diffusion equation. Thus, one can avoid using the Green's functions that take into account the deterministic force and for which only rarely analytical representations can be obtained.  
We will return to this point in section \ref{uniMolecular}.

\section{Theory}
\subsection{Smoluchowski equation}\label{smol_eq}
The Smoluchowski equation describes Brownian motion in terms of probability density functions $\rho(\mathbf{r}, t \vert \mathbf{r}_{0}, t_{0})$. 
The proposed algorithm is based on the possibility to describe an isolated pair of diffusing particles that may react with each other upon encounter as the diffusion of a point-like particle near a boundary \cite{Agmon:1990p10, BarSchieber:2002BS, vanZon:2005p401}.  
More precisely, one considers two spherical diffusing molecules $A$ and $B$. Their probability density function is described by the two-body Smoluchowski equation: 
\begin{equation}\label{2bodySmol}
\frac{\partial}{\partial t} \rho(\mathbf{r}_{A},  \mathbf{r}_{B}, t \vert \mathbf{r}_{A0},  \mathbf{r}_{B0}, t_{0})= (D_{A}\nabla_{A}^{2}+D_{B}\nabla_{B}^{2})\rho(\mathbf{r}_{A},  \mathbf{r}_{B}, t \vert \mathbf{r}_{A0},  \mathbf{r}_{B0}, t_{0})
\end{equation}
where $D_{A}, D_{B}$ denote the diffusion constants of molecule A and B, respectively.
 By transition to the coordinates
\begin{eqnarray}
\mathbf{R}&=&\frac{D_{B}}{D_{\text{eff}}} \mathbf{r}_{A}+\frac{D_{A}}{D_{\text{eff}}} \mathbf{r}_{B}\label{Rcoo} \\
\mathbf{r}&=&\mathbf{r}_{B}-\mathbf{r}_{A}\label{relCoo}
\end{eqnarray}
we obtain  
\begin{equation}
\frac{\partial}{\partial t} \rho(\mathbf{R},  \mathbf{r}, t \vert \mathbf{R},  \mathbf{r}, t_{0}) = (D_{R}\nabla_{\mathbf{R}}^{2} + D_{\text{eff}}\nabla_{\mathbf{r}}^{2})\rho(\mathbf{R},  \mathbf{r}, t \vert \mathbf{R},  \mathbf{r}, t_{0}),
\end{equation}
where 
\begin{eqnarray}
D_{\text{eff}} &=& D_{A}+D_{B} \\
D_{R} &=& \frac{D_{A}D_{B}}{D_{\text{eff}}}.
\end{eqnarray}
Evidently, the two-body Smoluchowski equation \eqref{2bodySmol} governs two independent random processes. Hence, the above equation for $\rho(\mathbf{R},  \mathbf{r}, t \vert \mathbf{R},  \mathbf{r}, t_{0})$ may be rewritten as one equation for $\mathbf{R}$ and one for the relative inter-particle vector $\mathbf{r}$
 \begin{eqnarray}
\frac{\partial}{\partial t}\rho(\mathbf{R}, t\vert \mathbf{R}_{0}, t_{0}) &=& D_{\text{R}}\nabla^{2}_{\mathbf{R}} \rho(\mathbf{R}, t\vert \mathbf{R}_{0}, t_{0})\label{center_eq}\\\
\frac{\partial}{\partial t}\rho(\mathbf{r}, t\vert \mathbf{r}_{0}, t_{0}) &=& D_{\text{eff}}\nabla^{2}_{\mathbf{r}} \rho(\mathbf{r}, t\vert \mathbf{r}_{0}, t_{0}), \quad \mathbf{r} \geq a_{\text{eff}}.\label{smol_bc}
\end{eqnarray}
$a_{\text{eff}}$ denotes the encounter radius and - in the absence of long range interaction potentials - would be given by the sum of the molecules' radii. Note that equation \eqref{smol_bc} describes a single particle diffusing with $D_{\text{eff}}$ that is excluded from a sphere with radius $a_{\text{eff}}$ located at the origin. We will return to this analogy in section \ref{Detection}.

Defining the probability flux by
\begin{equation}\label{flux}
 \mathbf{j}(\mathbf{r}, t\vert \mathbf{r}_{0}, t_{0}) := -D_{\text{eff}}\nabla_{\mathbf{r}} \rho(\mathbf{r}, t\vert \mathbf{r}_{0}, t_{0})
\end{equation}
the FP equation \eqref{smol_bc} may be written as continuity equation:
\begin{equation}\label{continuity}
\frac{\partial}{\partial t}\rho(\mathbf{r}, t\vert \mathbf{r}_{0}, t_{0})  + \nabla_{\mathbf{r}} \cdot \mathbf{j}(\mathbf{r}, t\vert \mathbf{r}_{0}, t_{0}) = 0, \quad \mathbf{r} \geq a_{\text{eff}}.
\end{equation}
The FP equations have to be completed by specifying boundary conditions for the conditional pdf $\rho(\mathbf{r}, t\vert \mathbf{r}_{0}, t_{0})$ and/or the probability flux \eqref{flux}.
Together with the following initial
\begin{equation}\label{initial_bc}
\rho(\mathbf{R}, t_{0}\vert \mathbf{R}_{0}, t_{0})=\delta(\mathbf{R}-\mathbf{R}_{0})
\end{equation}
and boundary condition
\begin{equation}\label{inf_bc}
\rho(\vert\mathbf{R}\vert\rightarrow\infty, t\vert \mathbf{R}_{0}, t_{0})=0
\end{equation}
equation \eqref{center_eq} is equivalent to the free-space diffusion equation with the familiar solution
\begin{equation}\label{gaussian}
\rho(\mathbf{R}, t\vert\mathbf{R}_{0}, t_{0})=\tfrac{1}{(4\pi D_{\text{R}}(t-t_{0}))^{3/2}}\,e^{-\frac{(\mathbf{R}-\mathbf{R}_{0})^{2}}{4D_{\text{R}}(t-t_{0})}},
\end{equation}
also known as the free-space Green's function. Henceforth, Green's functions will be denoted as $G(\mathbf{r}, t\vert \mathbf{r}_{0}, t_{0}) $.

The equation for the inter-particle vector $\mathbf{r}$ is only defined for $\mathbf{r} \geq a_{\text{eff}}$ and one has to impose a boundary condition for $\vert\mathbf{r}\vert=a_{\text{eff}}$ specifying the physics at the encounter distance. We will discuss the following cases \cite{Agmon:1990p10}:
\begin{itemize}
\item Absorbing boundary conditions: The molecules react instantaneously upon encounter.
\begin{equation}\label{g_abs_bc}
G(\vert \mathbf{r} \vert= a_{\text{eff}}, t\vert \mathbf{r}_{0}, t_{0}) = 0;
\end{equation}
\item Reflective boundary conditions: The molecules never react upon encounter.
\begin{equation}\label{g_ref_bc}
J(\vert\mathbf{r}\vert=a_{\text{eff}}, t \vert \mathbf{r}_{0}, t_{0})  = 0.
\end{equation}
Here we defined the total flux through a sphere by
$J(\mathbf{r}, t \vert \mathbf{r}_{0}, t_{0})=\omega_{d}\,r^{d -1}\mathbf{n}(\mathbf{r}) \cdot \mathbf{j}(\mathbf{r}, t \vert \mathbf{r}_{0}, t_{0})$.
$d$ denotes the number of space dimensions, $\omega_{d}:=2\pi^{d/2}/\Gamma(d/2)$ is the surface area of the unit sphere in $d$ dimensions and $\mathbf{n}(\mathbf{r})$ denotes the unit outward directed normal vector at a boundary point $\vert\mathbf{r}\vert = a_{\text{eff}}$.
\item Radiation boundary condition \cite{collins1949diffusion}: Upon encounter the molecules undergo a chemical reaction with a certain probability, otherwise they get reflected.
\begin{equation}\label{robin_bc}
 J(\vert\mathbf{r}\vert = a_{\text{eff}}, t \vert \mathbf{r}_{0}, t_{0}) =\kappa_{a}G(\vert\mathbf{r}\vert=a_{\text{eff}},t\vert\mathbf{r}_{0},t_{0}).
\end{equation}
The constant parameter $\kappa_{a}$ relating the flux of probability to the probability that the particles are in contact is referred to as intrinsic association rate at contact. It is distinct from the reaction rate constant $k_{on}$, which appears in macroscopic (mass action) rate equations and which contains additional contributions from the diffusive behavior of the particles. For later reference and to make our convention for $\kappa_{a}$ clear, we give the radiation boundary condition for 3D and 2D more explicitly
\begin{eqnarray}
4\pi a^{2}_{\text{eff}} D \frac{\partial}{\partial r}G(\mathbf{r}, t\vert\mathbf{r}_{0}, t_{0})\vert_{\vert\mathbf{r}\vert=a_{\text{eff}}} &=&\kappa^{3D}_{a}G(\vert\mathbf{r}\vert=a_{\text{eff}}, t\vert \mathbf{r}_{0}, t_{0})\nonumber\\
2\pi a_{\text{eff}} D \frac{\partial}{\partial r}G(\mathbf{r}, t\vert\mathbf{r}_{0}, t_{0})\vert_{\vert\mathbf{r}\vert=a_{\text{eff}}} &=&\kappa^{2D}_{a}G(\vert\mathbf{r}\vert=a_{\text{eff}}, t\vert \mathbf{r}_{0}, t_{0}).\nonumber
\end{eqnarray} 
\end{itemize}
Clearly, \eqref{g_abs_bc} and \eqref{g_ref_bc} are the limit of \eqref{robin_bc} when $\kappa_{a}\rightarrow \infty$ and $\kappa_{a} \rightarrow 0$, respectively.
The described boundary conditions render the associated solution nontrivial. Notwithstanding, analytical representations for the full solutions are known for all three cases in 3D and 2D \cite{carslaw1986conduction}. Their numerical approximation is discussed in section \ref{numApp}. In section \ref{Detection} it will be shown how they can be used to remedy the issue of underestimating the number of particle encounters during a naive BD simulation time step.

Other quantities of interest can be derived from the Green's function. The survival probability of a pair of molecules separated by $\mathbf{r}_{0}$ at time $t_{0}$ not to react and thus survive until at least time $t$ is defined by
\begin{equation}\label{SurvFull}
S_{\text{abs,rad}}(t\vert \mathbf{r}_{0}, t_{0}) = \int_{\vert\mathbf{r}\vert>a_{\text{eff}}}G_{\text{abs,rad}}(\mathbf{r}, t\vert \mathbf{r}_{0}, t_{0})d^{3}\mathbf{r}.
\end{equation}
Note that one has two different survival probabilities corresponding to the boundary conditions involving particle absorption. For purely reflective bcs, the survival probability is equal to one for all times.
In the cases of spherical symmetry that we consider here, the survival probability depends only on the radial Green's function $g(r,t\vert r_{0}, t_{0})$ 
\begin{equation}\label{SurvRad}
S_{\text{abs,rad}}(t\vert r_{0}, t_{0}) = \omega_{d}\int^{\infty}_{a_{\text{eff}}}g_{\text{abs,rad}}(r, t\vert r_{0}, t_{0})r^{d-1}dr.
\end{equation}
Thus, for all three boundary conditions the radial Green's functions can be obtained by 
\begin{equation}\label{radialGF}
\omega_{d} g(r, t\vert r_{0}, t_{0})=\int G(\mathbf{r}, t\vert \mathbf{r}_{0}, t_{0}) d^{d}\Omega
\end{equation}
where $d^{d}\Omega$ means the infinitesimal surface element of the unit sphere in $d$ dimensions.
\subsection{Reversible bimolecular reactions and collision detection via Green's functions}\label{Detection}
In this section we introduce the key components of the proposed simulation algorithm. In particular, we will discuss expressions for 
\begin{itemize}
\item the encounter probability that will permit collision detection, 
\item the propagator that will tell us how to (re-)sample the new positions of the molecules involved in encounters,
\item the reaction probability that allows us to determine the pairs of encountered molecules that subsequently undergo a chemical reaction,
\item describing unimolecular reactions, in particular back-reactions $C\rightarrow A+B$.  
\end{itemize}
To this end we again describe two-particle systems in terms of the diffusional behavior of a point-like particle around a sphere in 3D and 2D, but instead of a description based on the Fokker-Planck equation as in \ref{smol_eq}, we switch to a decription in terms of individual trajectories and adopt the stochastic differential equations' point of view. Indeed, we will emphasize the relationship between both approaches by recalling how the Green's functions of the initial and boundary value problem arise naturally in the trajectory picture without ever using a partial differential equation. Although most of the 1D examples can be found in textbooks \cite{harrison1985brownian}, we will present them here for two reasons. First, we think that the chosen representation makes the underlying physics evident and facilitates the derivation of the relevant expressions greatly. Second, by clarifying the role of the first-passage time, the relationship with event-driven approaches can be established, cp. \ref{appendix}.
\subsubsection{Encounter Probability}\label{encProb}
Consider a point-like particle with initial position $\mathbf{r}_{0}$ at time $t_{0}$ in the vicinity of an absorbing sphere $\partial \mathcal{S}$ and let $\mathcal{S}$ denote the region of space internally bounded by $\partial \mathcal{S}$. In a naive BD simulation, the particle's displacement $\mathbf{r} - \mathbf{r}_{0}$ in the next time step is sampled according to the free-space Langevin equation \eqref{oLangevin}, i.e. according to a Gaussian pdf, cp. \eqref{gaussian}.

After the simulation time step the particle may end up in the "forbidden" region $\mathcal{S}$, which necessarily - assuming continuity of the trajectory - means that the particle has hit the boundary somewhere in the course of the time step. In the case of an absorbing boundary this is equivalent to the occurrence of an reaction. However, even if the particle's position after the time step is outside the boundary region, one cannot conclude that there was no encounter during $\Delta t$. 
This problem was analyzed in reference \cite{BarSchieber:2002BS} and it was shown how to address it by computing an "encounter probability", given that the particle arrives at $\mathbf{r}$ after the time step and given that it started a $\mathbf{r}_{0}$.
We will rederive that expression in a way that will be useful for later considerations. 
To this end we consider (mathematical) Brownian motion $\mathbf{W}_{t}$, which is a stochastic process, i.e., for every $t\geq 0$, $\mathbf{W}_{t}(\cdot): \omega\in\Omega \rightarrow \mathbf{W}(t, \omega)$ is a random variable in the probability space $(\Omega, \mathcal{A}, \text{Prob})$. We refrain from giving a more precise definition of $\mathbf{W}_{t}$ and $(\Omega, \mathcal{A}, \text{Prob})$ that can be found elsewhere \cite{harrison1985brownian}. Note that at the level of the stochastic differential equation $\mathbf{W}_{t}$ is related to the physical stochastic trajectory $\mathbf{X}_{t}$ by $d\mathbf{X}_{t}=\sqrt(2D)d\mathbf{W}_{t}$.

For simplicity, let us consider a 1D Brownian motion whose trajectories start at a point $x_{0} = 0$ at time $t_{0} = 0$ without loss of generality. Furthermore, we assume the presence of an absorbing boundary at $x = a > x_{0}$. Therefore, the "allowed" region is $[-\infty, a)$. The result of every simulation step corresponds to the stochastic outcome of an experiment and we are interested in assigning a probability to the following events:
\begin{equation}
A :=  \lbrace \text{there was no encounter with the boundary for all } s \leq t \rbrace,
\end{equation} 
and
\begin{equation}
B := \lbrace X_{t} = x \vert X_{t_{0}} = x_{0}\rbrace.
\end{equation} 
We wish to find the conditional probability 
\begin{equation}\label{bayes}
\text{Prob}(A\vert B)
\end{equation}
starting from the Wiener probability distribution
\begin{eqnarray}\label{WienerMeasure}
F_{W}(x, t\vert 0,0)&:=&\text{Prob}\lbrace W_{t} \leq x \vert W_{t_{0}=0} = 0\rbrace \nonumber\\
&=&\frac{1}{\sqrt{2\pi t}}\int^{x}_{-\infty}e^{-y^{2}/2t}dy
\end{eqnarray}
Henceforth, for notational simplicity, we will often suppress the condition $\vert W_{t_{0}=0} = 0$.
 
In this context, the presence of boundary conditions can be incorporated by introducing additional random variables. 
More precisely, at the level of individual trajectories an absorbing boundary can be realized by terminating any trajectory as soon as it hits the boundary for the first time.
Hence, for absorbing boundaries the first-passage time $\tau_{a}$ plays a central role. It is defined by  \cite{ottinger1996stochastic, redner2001guide}
\begin{equation}
\tau_{a} = \text{inf}\lbrace t \in [0, t] \, : \, W_{t} \notin [-\infty, a[ \rbrace.
\end{equation}
Note that $W_{\tau_{a}} = a$.
With the help of the first passage time one may define a reflected Brownian motion 
\begin{equation}\label{refPath}
\tilde{W}_{t} = \left\{\begin{array}{lr}
 W_{t} &\mbox{ $t < \tau_{a}$} \\
 2W_{\tau_{a}} - W_{t}  &\mbox{ $t \geq \tau_{a}$}
\end{array}\right .
\end{equation}
  
We are now in a position to calculate the joint probability distribution function $\text{Prob}(W_{t} \leq x, \tau_{a} \leq t)$. Using \eqref{refPath}, $\lbrace \omega \vert W_{t} \geq 2a - x\rbrace \subset \lbrace \omega \vert  \tau_{a} \leq t \rbrace$, the reflection principle and \eqref{WienerMeasure}, one obtains
\begin{eqnarray}\label{PDFMaxBigger}
\text{Prob}(W_{t} \leq x, \tau_{a} \leq t) &=& \text{Prob}(\tilde{W}_{t} \geq 2a - x;  \tau_{a} \leq t) \nonumber \\
	&=& 1 -   \text{Prob}(W_{t} \leq 2a - x) \nonumber \\
	&=&\frac{1}{\sqrt{2\pi t}}\int^{\infty}_{2a-x}e^{-y^{2}/2t}dy
\end{eqnarray}
Because of \eqref{WienerMeasure} and 
\begin{equation}
\text{Prob}(W_{t} \leq x) = \text{Prob}(W_{t} \leq x, \tau_{a} \leq t) + \text{Prob}(W_{t} \leq x, \tau_{a} > t),
\end{equation}
it follows 
\begin{eqnarray}\label{stochGabs}
\text{Prob}(W_{t} \leq x, \tau_{a} > t ) &=& \text{Prob}(W_{t} \leq x) - \text{Prob}(W_{t} \leq x, \tau_{a} \leq t) \\
& = &\frac{1}{\sqrt{2\pi t}}\left(\int^{x}_{-\infty}e^{-y^{2}/2t}dy-\int^{\infty}_{2a - x }e^{-y^{2}/2t}dy\right).\nonumber
\end{eqnarray}
The joint probability distribution function \eqref{stochGabs} takes into account all Brownian paths that end at $x$ at $t$ and whose first-passage time is bigger than $t$, and thus never encountered the boundary during the time interval. The associated pdf is 
\begin{eqnarray}\label{2pabs}
f_{W}(x, t\vert 0, 0) &=& \frac{\partial}{\partial x}\text{Prob}(W_{t} \leq x, \tau_{a} > t )\nonumber\\
& = & \text{Prob}(W_{t}=x, \tau_{a} > t)\nonumber\\
&=&\frac{1}{\sqrt{2\pi t}}\left(e^{-x^{2}/2t}-e^{-(x - 2a)^{2}/2t}\right).
\end{eqnarray}
which is the Green's function of the initial and absorbing boundary problem of the 1D diffusion equation.
Importantly, this relation between the joint probability density function of the first passage time and the absorbing Green's function also holds in 2D and 3D.

Because \eqref{2pabs} is $\text{Prob}(A\cap B)$, we can now calculate \eqref{bayes}, the probability that the Brownian mover never touched the boundary within the time step, given that $W_{t} = x$, according to Bayes' formula, cp. \cite{BarSchieber:2002BS}
\begin{eqnarray}\label{bayes1D}
\text{Prob}(\tau_{a} >  t \, \vert\, W_{t} = x)&=&\frac{\text{Prob}(W_{t} = x; \tau_{a} > t)}{\text{Prob}(W_{t} = x)}=\nonumber\\
\frac{G_{\text{abs}}(x, t\vert x_{0}, t_{0})}{G_{\text{free}}(x,  t\vert x_{0}, t_{0})}&=:& 1-p_{\text{enc}}(x, t\vert x_{0}, t_{0})
\end{eqnarray}
\eqref{bayes1D} provides a way to detect an encounter of particles, even if after the simulation time step the particles have no overlap. Henceforth, we refer to $p_{\text{enc}}(x, t, x_{0}, t_{0})$ as encounter probability .

 Although we considered the 3D case for simplicity, the key relationship between first-passage time and the Green's function satisfying absorbing bcs does also hold in 2D and 3D and the whole line of reasoning can be extended to these cases. However, to this end one has to substitute in \eqref{bayes1D} the appropriate pdfs (Green's functions), which are known \cite{carslaw1986conduction}, cp. section \ref{numApp} \eqref{genericGF3D} and \eqref{genericGF}.

\subsubsection{Propagation}\label{prop}
For a completely diffusion limited reaction, the encounter probability is equivalent to the reaction probability: upon detecting an encounter the algorithm replaces the pair of particles by its reaction products. However, if a chemical reaction is not completely diffusion limited, a particle reacts only with a finite rate upon hitting the encounter surface and is otherwise reflected. Therefore, the particles that did encounter each other must not be propagated according to the free-space Green's function, but the propagation needs to take into account the reflective encounter, cp. \cite{Lamm:1983p12, Lamm:1984p4, North:1984, NorthCurv:1986, green1988simulation, pimblott1992stochastic, ottinger1996stochastic}.
In a naive BD simulation, reflecting bcs are frequently incorporated in an analogous manner to absorbing bcs: The terminal positions of only those particles that overlap after a time step $\Delta t$ are reset to the boundary $\mathbf{r}\in\partial \mathcal{S}$ \cite{ottinger1996stochastic}. This procedure introduces two errors: First, as in the case of absorbing boundary conditions, the number of actual reflections is underestimated. Second,  the resetting is only justified if the Brownian motion assumes a boundary position at the terminal time $t = t_{0} + \Delta t$, but never crossed the boundary during the time step. In general, however, it has crossed the boundary at an earlier time $t_{0}<t^{\prime} <  t_{0}+\Delta t$. 

To correct for the first error, we can employ the same strategy as in the aborbing boundary case: We recall that \eqref{bayes1D} gives the probability that the particle has hit the boundary which, for an absorbing bc, means that a reaction happened, but which means in the present context that a reflection occurred. To correct for the second error, instead of using the resetting method, we employ the Green's function satisfying reflecting boundary conditions:
\begin{equation}\label{green_reflect}
\mathbf{n}(\mathbf{r})\cdot \nabla_{\mathbf{r}}G_{\text{ref}}(\mathbf{r}, t_{0}+\Delta t | \mathbf{r}_{0}, t_{0})|_{\mathbf{r}\in\partial \mathcal{S}} = 0
\end{equation}

However, the new positions of the encountered particles have to be sampled according to 
\begin{equation}\label{passage_prob}
G_{\text{refl}}(\mathbf{r}, t_{0}+\Delta t | \mathbf{r}_{0}, t_{0})-G_{\text{abs}}(\mathbf{r}, t_{0}+\Delta t | \mathbf{r}_{0}, t_{0})
\end{equation}
instead of $G_{\text{refl}}(\mathbf{r}, t_{0}+\Delta t | \mathbf{r}_{0}, t_{0})$. This can be seen as follows: Consider in 1D the reflected Brownian motion 
\begin{equation}\label{refBP}
\bar{W}_{t} := \left\{\begin{array}{lr}
 W_{t} &\mbox{if $W_{t} \leq a $} \\
 2a - W_{t} &\mbox{if $W_{t} > a$}  
\end{array}\right .
\end{equation}
and the corresponding probability distribution $\text{Prob}(\bar{W}_{t} \leq x)$ for $ x < a$. The associated probability density function is the Green's function of the initial and reflective boundary problem of the FP equation.
Now, the probability distribution may be written as
\begin{equation}\label{stochGref}
\text{Prob}(\bar{W}_{t} \leq x) = \text{Prob}(\bar{W}_{t} \leq x, \tau_{a} \leq t) + \text{Prob}(W_{t} \leq x, \tau_{a} > t), 
\end{equation}
because $\lbrace \tau_{a} \leq t\rbrace$ and $\lbrace \tau_{a} > t\rbrace$ are mutually exclusive events and due to \eqref{refBP}. As discussed in section \ref{encProb}, the second term on the right hand side yields the Green's function with absorbing bcs, cp. \eqref{PDFMaxBigger} and \eqref{stochGabs}, taking into account those particular trajectories of the reflected Brownian motion that did not get reflected. But in the simulation one only resamples the terminal points of the particles that did undergo an reflection. Thus, one has to sample according to  $\text{Prob}(\bar{W}_{t} \leq x, \tau_{a} \leq t) $, but this is $G_{\text{ref}}(x, t_{0}+\Delta t | x_{0}, t_{0})-G_{\text{abs}}(x, t_{0}+\Delta t | x_{0}, t_{0}).$

These results can readily be extended to higher dimensions and thus we arrive at the following strategy to improve the accuracy of a naive BD simulation of reflecting particles: 
Sample the positions of all particles according to the unbounded standard process. If two particles overlap, they must have reflected each other and are resampled according to 
\eqref{passage_prob}
to avoid the error made by resetting the particles' positions at the encounter distance.
For all other pairs of particles we calculate the encounter probability using equation \eqref{bayes1D}. If there was a reflection, the new positions of the particles are resampled according to \eqref{passage_prob}.

\subsubsection{Bimolecular reaction probability}\label{reacProb}
In addition, we now assume that there is a finite probability that the particle has actually undergone a reaction. To take this into account the conditional probability that the particle has reacted upon reflection, given its initial position before the reflection is $\mathbf{r}_{0}$ and its position after the reflection event is $\mathbf{r}$, can be employed. Again, Bayes' formula can be used to find this conditional reaction probability, but the condition can no longer be described in terms of $G_{\text{free}}(\mathbf{r}, t_{0}+\Delta t | \mathbf{r}_{0}, t_{0})$ as in \eqref{bayes1D}, because one considers only the pairs that got reflected and these were propagated according to $G_{\text{refl}}(\mathbf{r}, t_{0}+\Delta t | \mathbf{r}_{0}, t_{0}) - G_{\text{abs}}(\mathbf{r}, t_{0}+\Delta t | \mathbf{r}_{0}, t_{0})$ instead of according to the free-space Green's function. Now, it only remains to find the numerator in Bayes' formula. To this end one may argue in a similar way as in \ref{prop}. Because only reflected pairs can react, they have to be described by $G_{\text{ref}}(\mathbf{r}, t_{0}+\Delta t | \mathbf{r}_{0}, t_{0})$, but as pointed out in \ref{prop}, the $G_{\text{ref}}$ has also contributions from the paths that did not lead to an encounter (and which can be subtracted by $G_{\text{abs}}$), but in the presence of partially absorbing bcs also contributions from paths that involved an reflection, but nevertheless did not react. These contributions are taken into account by $G_{\text{rad}}$. Thus, for the numerator in Bayes' formula one may write $G_{\text{ref}}(\mathbf{r}, t_{0}+\Delta t | \mathbf{r}_{0}, t_{0}) - G_{\text{rad}}(\mathbf{r}, t_{0}+\Delta t | \mathbf{r}_{0}, t_{0})$ and, in total, for the reaction probability one obtains according to Bayes' formula, cp. also \cite{Lamm:1983p12, Lamm:1984p4, North:1984, NorthCurv:1986, green1988simulation, pimblott1992stochastic}
\begin{equation}\label{psurv_radII}
p_{\text{reac}}(\mathbf{r}, \mathbf{r}_{0}, \Delta t ) =\frac{G_{\text{ref}}(\mathbf{r}, t_{0}+\Delta t | \mathbf{r}_{0}, t_{0}) - G_{\text{rad}}(\mathbf{r}, t_{0}+\Delta t | \mathbf{r}_{0}, t_{0})}{G_{\text{ref}}(\mathbf{r}, t_{0}+\Delta t | \mathbf{r}_{0}, t_{0}) - G_{\text{abs}}(\mathbf{r}, t_{0}+\Delta t | \mathbf{r}_{0}, t_{0})}
\end{equation} 

It is straightforward to see that the method for radiation boundary conditions reduces to the method for absorbing boundaries in the limit $\kappa_{a}\rightarrow\infty$. As $G_{\text{rad}}(\mathbf{r}, t_{0}+\Delta t | \mathbf{r}_{0}, t_{0})\underset{\kappa_{a}\rightarrow\infty}{\rightarrow}G_{\text{abs}}(\mathbf{r}, t_{0}+\Delta t | \mathbf{r}_{0}, t_{0})$ it follows from \eqref{psurv_radII} that in this limit all particles that would have been reflected (which means that they reached the encounter distance) will react. This is exactly the method used in the case of absorbing boundary conditions. On the other hand, in the limit $\kappa_{a}\rightarrow0$ one obtains $G_{\text{rad}}(\mathbf{r}, t_{0}+\Delta t | \mathbf{r}_{0}, t_{0}){\rightarrow}G_{\text{ref}}(\mathbf{r}, t_{0}+\Delta t | \mathbf{r}_{0}, t_{0})$ and for the encounter probability \eqref{psurv_radII} $p_{\text{reac}}(\mathbf{r}, \mathbf{r}_{0}, \Delta t ) \rightarrow 0$. Consequently, there will be no reactions and we recover the method for purely reflecting boundaries.

\subsubsection{Radial versions of $p_{\text{enc}}$ and $p_{\text{reac}}$}\label{radialPENC}
Finally, we would like to point out that instead of the full Green's functions including angle dependency, cp. \eqref{genericGF3D}, \eqref{genericGF}, their radial counterparts \eqref{radialGF} may be used for the calculation of $p_{\text{enc}}$ and $p_{\text{reac}}$. At least in 3D this offers the advantage that one can avoid the numerical approximation of an integral, cp. section \ref{gaNum}. In 2D the situation is different, but for small times we will derive an alternative expansion that does not involve an numerical integration, cp. \ref{smallTimeSec}. Note that even if the radial versions of $p_{\text{enc}}$ and $p_{\text{reac}}$ are used, the particles are propagated according to the full propagator \eqref{passage_prob}.    
\subsubsection{Unimolecular reactions}\label{uniMolecular}
The considerations made in the previous sections can be applied to bimolecular reactions.  To include also the decay of single particles $A\rightarrow reaction products$ we assume that this reaction can be described as a Poisson process.
The probability that the next reaction occurs between $t$ and $t+dt$ is given by
\begin{equation}\label{single_rate}
q(t\vert t_{0})= k_{d}e^{-k_{d}(t-t_{0})}dt,
\end{equation}
with $k_{d}$ being the particle's decay rate.
To obtain the single molecule reaction probability that there was a reaction during the small but finite (i.e. not infinitesimal) time step $\Delta t$, \eqref{single_rate} has to be integrated:
\begin{equation}\label{rad_prob}
p^{\text{uni}}_{\text{reac}}(\Delta t) = \int^{t_{0}+\Delta t}_{t_{0}} q(t\vert t_{0}) dt =1-e^{-k_{d}\Delta t}.
\end{equation}
Typically, one is interested in the unimolecular reaction $C \rightarrow A+B$. Therefore the question arises how the distance between $A$ and $B$ after the time step should be chosen. To answer this we note that when $C$ decays, the molecules $A$ and $B$ are at contact $\vert\mathbf{r }\vert = a_{\text{eff}}$ and hence obviously encountered each other. We can thus apply the strategy adopted in the section about reflective boundary conditions. More precisely, we propagate the molecules according to \eqref{passage_prob} with $\vert\mathbf{r}_{0}\vert = a_{\text{eff}}$. Finally, \eqref{psurv_radII} is used to decide if the molecules escape or recombine.

We conclude this section by pointing out that the methods described in this section may also be applied in the presence of a deterministic potential, without requiring Green's function of the full Smoluchoswki equation, which can rarely analytically represented anyway. As described in \cite{BarSchieber:2002BS}, also in this case the Green's functions of the pure diffusion equation can be used by separating the contributions of the deterministic and the random force. More precisely, within one time step, first, the particles' displacements are calculated according to the deterministic forces only and subsequently sampled according to Brownian motion. The encounter and reaction probabilities are then calculated by taking into account only starting and terminal position of the Brownian motion, instead of starting and terminal point of the combined motion.  
\section{Simulation algorithm}\label{simAlg}
In the previous sections we have shown how Green's functions can be used to improve the naive BD simulation of the diffusive behavior of particles in the vicinity of reactive boundaries. 
The encounter and reaction probabilities,  \eqref{bayes1D} and \eqref{psurv_radII}, respectively, play a crucial role. They permit to compute the probability that there was a reaction during a time step, even if the particles have no overlap after the time step. In this way, underestimating the number of actual reactions can be avoided. As  a consequence, the convergence behavior can be improved and the time step can be chosen bigger than the one used in a naive BD simulation. 
The simulation algorithm consists of the following steps:
\begin{enumerate}
\item Propagate all particles' positions according to the solution of the free-space Smoluchowski equation, i.e.  for the displacements one has $\mathbf{X}_{t_{0}+\Delta t}-\mathbf{X}_{t_{0}} = \sqrt{2D\Delta t}\mathbf{N}(0,1)$, where $\mathbf{N}(0,1)$ denotes the unit normal random variable with vanishing mean and variance equal one.
\item  Check for all overlapping particles if the involved species pairs imply absorbing, reflecting or radiation boundary conditions.
\begin{enumerate}
\item In case of absorbing bcs, the particles underwent a reaction and are replaced by the reaction products.
\item In case of reflecting boundaries, the particles' position is reset by the following two steps. First, sample the $\mathbf{R}$ coordinate \eqref{Rcoo} according to \eqref{gaussian}. Second, the relative distance vector \eqref{relCoo} is sampled according to $G_{\text{refl}}(\mathbf{r}, t_{0}+\Delta t | \mathbf{r}_{0}, t_{0})-G_{\text{abs}}(\mathbf{r}, t_{0}+\Delta t | \mathbf{r}_{0}, t_{0})$ \eqref{passage_prob}.
\item In case of radiation boundary conditions, proceed in the same way as in the case of reflecting boundary conditions. In addition, calculate \eqref{psurv_radII}, where $\mathbf{r}$ is given by the new, according to \eqref{passage_prob} resampled, positions. Draw a uniform random number $\xi$. If $\xi<p_{\text{reac}}$, there was a reaction. Replace the pair of particles by the corresponding reaction products.    
\end{enumerate}
\item Calculate for all pairs of particles that do not overlap the encounter probability \eqref{bayes1D}. Draw a uniform random number $\xi$. If $\xi<p_{\text{enc}}$, there was an encounter.
Apply to all in this way found encounter pairs the steps 2 (a), (b), (c).
\item Calculate for each molecule that can undergo a unimolecular reaction the associated survival probabilty \eqref{rad_prob}. Draw a uniform random number $\xi$. If $\xi<p^{\text{uni}}_{\text{reac}}$, the molecule decayed and is replaced by its reaction products. Propagate the reaction products and check for escape or recombination as described in section \ref{prop}, \ref{reacProb}.
\item Increase the system time by $\Delta t$. Repeat steps 1., 2., 3, 4.
\end{enumerate}
\subsubsection*{Details}
\begin{itemize}
\item In step (3a), a particular molecule might possess several neighbors which fulfill the encounter criterion $\xi<p_{\text{enc}}$.  In this way the algorithm generates a list of encounter candidates for a particular molecule. To decide which of the candidate molecules has actually been involved in the encounter process one can proceed in a way similar to how the Gillespie algorithm decides which reaction channel is going to "fire" \cite{gillespie1977exact}:
More precisely, if there are $N_{\text{cand}}$ neighbors which fulfill the encounter criteria, one can apply the following algorithm
\begin{itemize}
\item Calculate the sum of the (already in step 3 calculated survival probabilities $p_{\text{enc}}$) of all candidate molecules.
\begin{equation}
p_{\text{tot}} = \sum^{N_{\text{cand}}}_{i=1} p_{\text{enc,i}}
\end{equation}
\item Calculate the ratios $\frac{\sum^{j}_{i=1}p_{\text{enc,i}}}{p_{\text{tot}}}, j = 1,\ldots N_{\text{cand}}$. It follows 
\begin{equation}
0 < \frac{\sum^{1}_{i=1}p_{\text{enc,i}}}{p_{\text{tot}}}< \frac{\sum^{2}_{i=1}p_{\text{enc,i}}}{p_{\text{tot}}}\ldots< \frac{\sum^{N_{\text{cand}}}_{i=1}p_{\text{enc,i}}}{p_{\text{tot}}}=1
\end{equation}
\item Sample a uniform random number $\xi$. Find the $j$ which satisfies:
\begin{equation}
\frac{\sum^{j}_{i}p_{\text{enc,i}}}{p_{\text{tot}}}<\xi<\frac{\sum^{j+1}_{i}p_{\text{enc,i}}}{p_{\text{tot}}}
\end{equation} 
\item Pick the $j+1$ molecule as "encounter" molecule  and continue with the algorithm starting from 3 (a). 
\end{itemize}  

\item Propagation by  $G_{\text{refl}}(\mathbf{r}, t_{0}+\Delta t | \mathbf{r}_{0}, t_{0})-G_{\text{abs}}(\mathbf{r}, t_{0}+\Delta t | \mathbf{r}_{0}, t_{0})$ means the following:
\begin{itemize}
\item First, the radial propagator \\$\omega_{d} r^{d-1}\left(g_{\text{refl}}(r, t_{0}+\Delta t | r_{0}, t_{0}) - g_{\text{abs}}(r, t_{0}+\Delta t | r_{0}, t_{0})\right)$, \\which gives the probability of finding the radial coordinate in the interval $[r, r+dr[$ at time $t_{0}+\Delta t$, is used to sample a new radial coordinate.
\item Then,  $\mathcal{J} \big(G_{\text{refl}}(r, \Theta, t_{0}+\Delta t | r_{0}, \Theta_{0},  t_{0})-G_{\text{abs}}(r, \Theta, t_{0}+\Delta t | r_{0}, \Theta_{0}, t_{0})\big)$, which yields the probability of finding the angle between $\mathbf{r}$ and $\mathbf{r}_{0}$ inside the interval $[\Theta, \Theta + d\Theta[$, given that the radial coordinate is $r$ at $t$, is used to draw a new angle. The factor $\mathcal{J}$ is defined as $\mathcal{J}:=2\pi \sin(\Theta)r^{2}$ in 3D and as $\mathcal{J}:=r$ in 2D, respectively.
\end{itemize}

\item To decrease computational cost one can introduce a cut-off radius $r_{\text{cut-off}}$ such that only neighboring particles within that radius are tested for encounters according to described procedure. Following \cite{BarSchieber:2002BS} the cut-off radius might be determined by requiring that the survival probability \eqref{SurvFull}
satisfies
\begin{equation}
S(t_{0} + \Delta t \vert \mathbf{r}_{\text{cut-off}}, t_{0}) \overset{!}{=} 10^{-3}.
\end{equation}
for the chosen simulation time step $\Delta t$. Depending on the type of the bc that models the interaction between the considered species pair, one uses either $S_{\text{abs}}$ or $S_{\text{rad}}$.
\end{itemize}

Finally, we would like to emphasize again that the described algorithm is not exact, but relies on approximations, as is obvious in the case of several encounter neighbors described above. The accuracy will be analyzed in detail in a forthcoming publication.

%%%%%%%%%%%%%%%%%%%%%%%%%%%%%%%%%%%%%%%%%%%%%%%%%%%%%%%%%%%%%%%%%%%%%%%%%%%%%%%%%%%%%%%%%
\section{Numerical approximation of the Green's functions}\label{numApp}
\subsection{General aspects}\label{gaNum}
The crucial ingredient of the suggested algorithm is the use of the Green's functions, which permit to increase the simulation time step. On the other hand, the calculation of the Green's functions is the most costly part of the algorithm. Hence, an efficient calculation of the Green's functions is required.

The analytical representation for the Green's function describing the diffusion of a point particle around a partially absorbing sphere is known \cite{carslaw1986conduction}. In a scaled form, which is suitable for numerical approximations, it is given in 3D by

\begin{eqnarray}\label{genericGF3D}
G_{\text{rad}}(R, \Theta, \tau \vert R_{0}) &=& 
 \tfrac{1}{4\pi a^{3} \sqrt{R R_{0}}}\sum^{\infty}_{n = 0}(2n+1)P_{n}(\cos(\Theta))\times\nonumber\\
&&\int^{\infty}_{0}e^{-\tau x^{2}} F_{n+1/2}(R, x) F_{n+1/2}(R_{0}, x)x\,dx.
\end{eqnarray}
The functions $F_{\nu}$, where $\nu = n+1/2$, are defined by
\begin{equation}
F_{\nu}(R, x)=\tfrac{(2 \tilde{h}^{\text{3D}}  + 1)[J_{\nu}(Rx)Y_{\nu}(x) - Y_{\nu}(Rx)J_{\nu}(x)] - 2 x [J_{\nu}(Rx)Y^{\prime}_{\nu}(x) - Y_{\nu}(Rx)J^{\prime}_{\nu}(x)]}{\lbrace [ (2 \tilde{h}^{\text{3D}}  + 1)J_{\nu}(x) - 2 x J^{\prime}_{\nu}(x)]^{2} + [(2 \tilde{h}^{\text{3D}} + 1)Y_{\nu}(x) - 2 x Y^{\prime}_{\nu}(x)]^{2}\rbrace^{1/2}}. 
\end{equation}
 Here, $R=r/a, R_{0}=r_{0}/a$  denote the dimensionless relative radial coordinates after and before the time step, respectively, and $\Theta$ denotes the angle between the corresponding relative position vectors. Furthermore,  $\tau = Dt/a^{2}$ is the dimensionless time and 
\begin{equation}
\tilde{h}^{\text{3D}} := h^{\text{3D}}a := \frac{\kappa^{\text{3D}}_{a}}{4\pi a D}
\end{equation}
might be thought of as a dimensionless reaction constant. 
$J_{n+1/2}, Y_{n+1/2}$ are the fractional Bessel functions of first and second kind \cite{abramowitz1964handbook}, respectively, and $P_{n}$ denote the Legendre ploynomials of order $n$ \cite{abramowitz1964handbook}.
For $\tilde{h}^{\text{3D}}\rightarrow\infty$ and $\tilde{h}^{\text{3D}}\rightarrow 0$, \eqref{genericGF3D} reduces to the Green's functions satisfying absorbing and reflecting boundary conditions, respectively.

In 2D the corresponding Green's function is
\begin{equation}\label{genericGF}
G_{\text{rad}}(R, \theta, t\vert R_{0}) = \frac{1}{2\pi a^{2}}  \sum^{\infty}_{n=-\infty} \cos(n\theta) \int^{\infty}_{0} e^{-\tau x^{2}}  C_{n}(R, x)C_{n}(R_{0}, x) x dx
\end{equation}
 where the functions  $C_{n}(R, x)$ are defined by
\begin{equation}\label{fRad}
C_{n}(R, x) = \frac{ J_{n}(Rx)[xY^{\prime}_{n}(x) - \tilde{h}^{\text{2D}} Y_{n}(x)] - Y_{n}(Rx)[xJ^{\prime}_{n}(x)-\tilde{h}^{\text{2D}}J_{n}(x)]}{([xJ^{\prime}_{n}(x) - \tilde{h}^{2D} J_{n}(x)]^{2} + [xY^{\prime}_{n}(x) - \tilde{h}^{2D} Y_{n}(x)]^{2})^{1/2}} ,
\end{equation}
and
\begin{equation}\label{2Dh}
\tilde{h}^{\text{2D}} := h^{\text{2D}}a := \frac{\kappa^{\text{2D}}_{a}}{2\pi  D}.
\end{equation}
Note that in \eqref{genericGF} the Bessel functions of first and second kind are of integer order \cite{abramowitz1964handbook}.
Again, one can obtain the Green's function for absorbing and reflecting bcs from \eqref{genericGF} by taking $\tilde{h}^{\text{2D}}\rightarrow\infty$ and $\tilde{h}^{\text{2D}}\rightarrow 0$.

The 3D and 2D radial Green's functions are given, up to a factor, by the zeroth term of the expansions \eqref{genericGF3D} and \eqref{genericGF}, respectively. Now, a special property of fractional Bessel functions is that they can be expressed in terms of elementary functions \cite[eq. 10.1.11, 10.1.12]{abramowitz1964handbook}
\begin{eqnarray}
J_{1/2}(x) &=& \sqrt{\frac{2}{\pi x}} \sin(x)\\
Y_{1/2}(x) &=& -\sqrt{\frac{2}{\pi x}} \cos(x)
\end{eqnarray}
Thus, the integral that involves the function $F_{1/2}$ and that yields the radial Green's functions can actually be solved explicitly for all three boundary conditions \cite{carslaw1986conduction}. Here we give them again a scaled form
\begin{eqnarray}
g_{\text{abs}}(R, \tau\vert R_{0})  = \tfrac{1}{ 8\pi a^{3} R R_{0}}\tfrac{1}{\sqrt{\pi \tau}}\left(\exp\left[-\tfrac{(R-R_{0})^{2}}{4\tau}\right] - \exp\left[-\tfrac{(R + R_{0} - 2)^{2}}{4\tau}\right]\right), 
\end{eqnarray}
and
\begin{eqnarray}
\lefteqn{g_{\text{rad}}(R, \tau\vert R_{0}) =}\nonumber\\
& &  \tfrac{1}{ 8\pi a^{3} R R_{0}}\tfrac{1}{\sqrt{\pi \tau}}\bigg[\exp\left[-\tfrac{(R-R_{0})^{2}}{4\tau}\right] + \exp\left[-\tfrac{(R + R_{0} - 2)^{2}}{4\tau}\right] + \nonumber\\
& & -\kappa \sqrt{4\pi\tau}\exp\left(\kappa^{2}\tau + (R+R_{0} - 2)\kappa\right)\text{erfc}\left( \kappa\sqrt{\tau} + \tfrac{(R+R_{0} - 2)}{2\sqrt{\tau}}\right)\bigg].
\end{eqnarray}
For convenience, we put  
\begin{equation}
\kappa := \frac{\kappa^{\text{3D}}_{a} + 4\pi aD}{4\pi aD}.
\end{equation}
$g_{\text{ref}}$ is given by the same expression as $g_{\text{rad}}$ with $\kappa^{3D}_{a}=0$.

By contrast, in the 2D case the corresponding integrals involve Bessel functions of integer order for which no expression in terms of elementary functions is known and hence, the calculation of the 2D radial Green's functions requires a numerical integration. As a consequence, the 2D simulation is more costly in this regard than a 3D simulation, because, as described in section \ref{simAlg}, the radial Green's functions are used by a random number generator to sample new positions. Moreover, the radial versions of $p_{\text{enc}}$ and $p_{\text{reac}}$ mentioned in \ref{radialPENC} that permit faster simulations in 3D, loose this advantage in the 2D case. In particular for smaller $\tau$ this leads to a substantially increased computational cost when compared to the 3D case.
To address this problem we will derive in section \ref{smallTimeSec} a small time expansion for the radial Green's functions satisfying absorbing, reflecting and radiation boundary conditions, respectively. 

The analytical representations \eqref{genericGF3D} and \eqref{genericGF} can be numerically approximated by the use of three-term recurrence relations \cite{deuflhard2003numerical}, which are satisfied by the functions $P_{n}, \cos(n\theta), J_{\nu}, Y_{\nu}$: For $P_{n}$ one has \cite[eq. 8.5.3]{abramowitz1964handbook}
\begin{equation}
(n+1)P_{n+1}(x)=(2n+1)xP_{n}(x)-nP_{n-1}(x),
\end{equation} 
where $P_{0}(x) = 1, P_{1}(x) = x.$ 
Both the Bessel functions $J_{\nu}(x), Y_{\nu}(x)$ satisfy the same three-term recurrence relations, but due to the fact that for $J_{\nu}(x)$ one has to iterate down it is more convenient to write them as \cite[eq. 9.1.27]{abramowitz1964handbook}
\begin{eqnarray}\label{3termBessel}
J_{\nu}(x) = \dfrac{2(\nu+1)}{x} J_{\nu+1}(x) - J_{\nu+2}(x)\\
Y_{\nu+2}(x) = \dfrac{2(\nu+1)}{x} Y_{\nu+1}(x) - Y_{\nu}(x)
\end{eqnarray}
Note that \eqref{3termBessel} can be used for both Bessel functions of integer and fractional order, cp \cite{BarSchieber:2002BS}.
Finally, the $c_{n}:=\cos(n\theta)$ satisfy
\begin{equation}
c_{n+2} = 2\cos(x)c_{n+1} - c_{n},
\end{equation}
where $c_{0} = 1, c_{1} = \cos(x)$.
For the integration procedure we chose a Gaussian integration rule. The advantage of this method is that not only the weights and abscissas $\lbrace x_{i}\rbrace$, but also the values of the Bessel functions at the abscissas (but not at $Rx_{i}, R_{0}x_{i}$) can be calculated in advance, i.e. before the actual simulation starts.  
\subsection{Small time expansion of 2D Green's functions}\label{smallTimeSec}
For small times, i.e. in terms of the dimensionless time $\tau:= D t / a^{2} \leq 0.01$, the representation of the Green's functions given above becomes cumbersome. The involved integrals become increasingly difficult to solve numerically due to the oscillatory character of the Bessel functions which become less and less dampened by the exponential factor.
Furthermore, as already discussed above, in 2D the integrals yielding the radial Green's functions cannot be solved analytically, in contrast to the 3D case. As a consequence, the 2D case is actually more cumbersome than the 3D case in the context of the suggested simulation algorithm (actually in the context of any algorithm based on the 2D Green's functions).

To address this problem we will show in the following how one can obtain an alternative expression for the 2D radial Greens's functions in terms of a small time expansion and thus, effectively, solve the integral.

\cite{carslaw1986conduction} demonstrates how the Laplace transform technique can be used to find small time expansions for a number of solutions to the heat equation. To the best of our knowlege, the specific small time expansions we are deriving here have not been discussed yet.
We make the following ansatz for the Laplace transform of the radial Green's function that satisfy certain boundary conditions \cite{carslaw1986conduction}
\begin{equation}\label{laplaceAnsatz}
\tilde{g}(r, q \vert r_{0}) = \tilde{g}_{\text{free}}(r, q \vert r_{0}) +\tilde{g}_{\text{bc}}(r, q \vert r_{0}) 
\end{equation}
Here 
\begin{equation}\label{laplaceFree}
\tilde{g}_{\text{free}}(r, q \vert r_{0}) = \frac{1}{2\pi D}\left\{\begin{array}{lr}
 I_{0}(qr_{0}) K_{0}(qr)&\mbox{$ r  >  r_{0} $} \\
 I_{0}(qr) K_{0}(qr_{0}) &\mbox{$r  <  r_{0}$}  
\end{array}\right .
\end{equation}
is the Laplace transform of the radial free-space Green's function. $q$ is defined by $ q:=\sqrt{\tfrac{p}{D}} $, where $p$ denotes the Laplace domain variable. The part $\tilde{g}_{\text{bc}}$ that takes into account the boundary condition is a solution to the Laplace transformed 2D diffusion equation 
\begin{equation}\label{homDG}
\frac{d^{2}\tilde{g}_{\text{bc}}}{dr^{2}} + \frac{1}{r}\frac{d\tilde{g}_{\text{bc}}}{dr} - q^{2} \tilde{g}_{\text{bc}} = 0.
\end{equation}

The general solution to \eqref{homDG} is $AK_{0}(qr) + BI_{0}(qr)$, where $I_{0}(x), K_{0}(x)$ refer to the modified Bessel functions of first and second kind, respectively, and of order zero \cite{abramowitz1964handbook}.
Because we require $\lim_{x\rightarrow \infty}\tilde{g}_{\text{bc}}\rightarrow 0 $, and $\lim_{x\rightarrow\infty}I_{0}(x)\rightarrow \infty$, $B$ has to vanish and hence, $\tilde{g}_{\text{bc}}(r, q\vert r_{0}) = A K_{0}(qr)$.
$A$ is determined by the requirement that the complete Green's function $\tilde{g}(r, q\vert r_{0})$ \eqref{laplaceAnsatz} satisfies either absorbing, reflective or radiation boundary conditions, respectively
\begin{eqnarray}
\tilde{g}(r, q\vert r_{0})\vert_{r=a} &=& 0 \label{absBC}\\
\frac{d\tilde{g}(r, q\vert r_{0})}{dr}\vert_{r=a} &=& 0 \label{reflectiveBC}\\
\frac{d\tilde{g}(r, q\vert r_{0})}{dr}\vert_{r=a} &=& h^{\text{2D}} \tilde{g}(r, q\vert r_{0})\vert_{r=a}. \label{radBC}
\end{eqnarray}
$h^{\text{2D}}$ has been defined in \eqref{2Dh}. In the following we will use $h$ instead of $h^{\text{2D}}$.
We switch to dimensionless variables $\tilde{q}:=qa,\, \tilde{h}:= ha, \, R:=r/a,\, R_{0}:=r_{0}/a$ and obtain from \eqref{laplaceAnsatz}, \eqref{laplaceFree}, \eqref{absBC}, \eqref{reflectiveBC} and \eqref{radBC} for the corresponding  $\tilde{g}_{\text{bc}}(R, \tilde{q}\vert R_{0})$ components
\begin{eqnarray}
\tilde{g}_{\text{abs}}(R, \tilde{q}\vert R_{0})  &=& -\frac{1}{2\pi D} \frac{I_{0}(\tilde{q})}{K_{0}(\tilde{q})}K_{0}(\tilde{q}R_{0})K_{0}(\tilde{q}R).\label{absPert}\\ 
\tilde{g}_{\text{ref}}(R, \tilde{q}\vert R_{0}) &=& -\frac{1}{2\pi D} \frac{I^{\prime}_{0}(\tilde{q})}{K^{\prime}_{0}(\tilde{q})}K_{0}(\tilde{q}R_{0})K_{0}(\tilde{q}R).\label{refPert} \\ 
\tilde{g}_{\text{rad}}(R, \tilde{q}\vert R_{0})  &=& -\frac{1}{2\pi D} \frac{\frac{\tilde{h}}{\tilde{q}}I_{0}(\tilde{q})-I^{\prime}_{0}(\tilde{q})}{\frac{\tilde{h}}{\tilde{q}}K_{0}(\tilde{q})-K^{\prime}_{0}(\tilde{q})}K_{0}(\tilde{q}R_{0})K_{0}(\tilde{q}R).\label{radPert}
\end{eqnarray}
Instead of applying the inversion theorem for the Laplace transformation to the complete solution of the boundary problem given above \eqref{laplaceAnsatz},
we make the following detour. First, we note that in almost all for the algorithm required expressions the free-space part cancels out, so we focus on the boundary components. 
Second, for small times we are interested in obtaining expansions in powers of $p^{-1}$, or equivalently, $q^{-1}$.
To this end we exploit the asymptotic expansions of the modified Bessel functions for large arguments \cite{abramowitz1964handbook}
\begin{eqnarray}
I_{\nu}(x) \sim \frac{e^{x}}{\sqrt{2\pi x}}\,\sum_{k=0}^{\infty}(-1)^{k}\frac{a_{k}(\nu)}{x^{k}}\\
K_{\nu}(x) \sim \sqrt{\frac{\pi}{2 x}} \,e^{-x}\,\sum_{k=0}^{\infty}\frac{a_{k}(\nu)}{x^{k}}
\end{eqnarray}
where the coefficients are
\begin{equation}\label{coeffDef}
a_{k}(\nu)=\frac{(4\nu^{2}-1^{2})(4\nu^{2}-3^{2})\cdots(4\nu^{2}-(2k-1)^{2})}{k!8^{k}}
\end{equation}
Using these asymptotic expansions and
\begin{eqnarray}
I^{\prime}_{0}(x) &=& I_{1}(x), \\
K^{\prime}_{0}(x) &=& -K_{1}(x)
\end{eqnarray}
one can transform the exact expressions \eqref{absPert}, \eqref{refPert} and \eqref{radPert}, according to the rules for multiplication and division of asymptotic series \cite{bender1978advanced}, to the following asymptotic expansion in powers of  $q^{-1}$.
\begin{eqnarray}\label{expansionLaplace}
\tilde{g}_{\text{abs}} &\sim& -\frac{1}{4\pi D \sqrt{RR_{0}}} \frac{e^{-\tilde{q}(R+R_{0}-2)}}{\tilde{q}}\sum^{\infty}_{k=0}\frac{\gamma^{\text{abs}}_{k}(R, R_{0})}{\tilde{q}^{k}}, \\
\tilde{g}_{\text{ref, rad}} &\sim& \frac{1}{4\pi D \sqrt{RR_{0}}} \frac{e^{-\tilde{q}(R+R_{0}-2)}}{\tilde{q}}\sum^{\infty}_{k=0}\frac{\gamma^{\text{ref, rad}}_{k}(R, R_{0})}{\tilde{q}^{k}}.
\end{eqnarray}
Note that the different boundary conditions lead to the same generic form of their asymptotic expansions, which only differ with regard to sign and the expansion coefficients $\gamma_{k}$. 
For convenience, we give the first two expansion coefficients ($\gamma^{\text{abs}}_{0} = \gamma^{\text{ref}}_{0} =\gamma^{\text{rad}}_{0}=1)$
\begin{eqnarray}
\gamma^{\text{abs}}_{1} &=& \frac{2RR_{0} - (R + R_{0})}{8RR_{0}}\nonumber\\
\gamma^{\text{abs}}_{2} &=& \frac{9(R^{2}+R^{2}_{0}) + 2RR_{0} - 4RR_{0}(R + R_{0})+4R^{2}R^{2}_{0}}{128R^{2}R^{2}_{0}}\nonumber\\
\gamma^{\text{rad}}_{1} &=& -\frac{RR_{0}(6+16\tilde{h}) + R + R_{0}}{8RR_{0}}\nonumber\\
\gamma^{\text{rad}}_{2} &=& \frac{9(R^{2}+R^{2}_{0}) + 2RR_{0} + RR_{0}(R + R_{0})(12+32\tilde{h})+R^{2}R^{2}_{0}(36+192\tilde{h} + 256\tilde{h}^{2})}{128R^{2}R^{2}_{0}}\nonumber.
\end{eqnarray}
$\gamma^{\text{ref}}_{1}, \gamma^{\text{ref}}_{2}$ can be obtained from $\gamma^{\text{rad}}_{1}, \gamma^{\text{rad}}_{2}$ for $\tilde{h} = 0$. 
We see that the explicit form of the coeffcients becomes quickly cumbersome, so that in a simulation all the required coefficients are calculated by iterative use of \eqref{coeffDef}.

To find the small time expansion in the time domain we use the Laplace transforms \cite{carslaw1986conduction}
\begin{eqnarray}
\frac{e^{-qx}}{q} &\rightarrow& \left(\frac{D}{\pi t}\right)^{1/2}e^{-\frac{x^{2}}{4Dt}}\\
\frac{e^{-qx}}{p^{1+n/2}} &\rightarrow& (4t)^{n/2}\text{i}^{\text{n}}\text{erfc}(\frac{x}{2\sqrt{Dt}})
\end{eqnarray}
The functions $\text{i}^{\text{n}}\text{erfc}(x)$ are defined by
\begin{equation}\label{iErf}
\text{i}^{\text{n}}\text{erfc}(x) := \int^{\infty}_{x} \text{i}^{\text{n-1}}\text{erfc}(\xi) d\xi
\end{equation}
and 
\begin{equation}
\text{i}^{0}\text{erfc}(x) := \text{erfc}(x), 
\end{equation}
cp.\cite{carslaw1986conduction}.
Moreover, 
\begin{equation}
\text{i}^{1}\text{erfc}(x) := \text{i}\text{erfc}(x) := \frac{1}{\sqrt{\pi}}e^{-x^{2}} - x\,\text{erfc}(x).
\end{equation}
None of the integrals in \eqref{iErf} have to be calculated, because the $\text{i}^{\text{n}}\text{erfc}(x)$ functions satisfy the recursion relation 
\begin{equation}
2n \,\text{i}^{\text{n}}\text{erfc}(x) = \text{i}^{\text{n-2}}\text{erfc}(x) - 2x\, \text{i}^{\text{n-1}}\text{erfc}(x)
\end{equation}
that allows to calculate the \eqref{iErf} swiftly.
In this way, we finally arrive at the expressions for small times in the time domain
\begin{multline}\label{smallTEabs}
g_{\text{abs}}(R,\tau\vert R_{0}) \sim -\frac{1}{4\pi a^{2} \sqrt{\tau RR_{0} } }\bigg[\pi^{-1/2}  e^{-\frac{(R+R_{0} - 2)^{2}}{4\tau}} + \\
\frac{1}{2}\sum^{\infty}_{n=1}\gamma^{\text{abs}}_{n}(R, R_{0}) \,\text{i}^{\text{n-1}}\text{erfc}\left(\frac{R+R_{0}-2}{2\sqrt{\tau}}\right)\, (2\sqrt{\tau})^{n}\bigg]
\end{multline}
and
\begin{multline}\label{smallTEabsII}
g_{\text{ref, rad}}(R,\tau\vert R_{0}) \sim \frac{1}{4\pi a^{2} \sqrt{\tau RR_{0} } }\bigg[\pi^{-1/2}  e^{-\frac{(R+R_{0} - 2)^{2}}{4\tau}} + \\
\frac{1}{2}\sum^{\infty}_{n=1}\gamma^{\text{ref, rad}}_{n}(R, R_{0}) \,\text{i}^{\text{n-1}}\text{erfc}\left(\frac{R+R_{0}-2}{2\sqrt{\tau}}\right)\, (2\sqrt{\tau})^{n}\bigg]
\end{multline}
Again, the differences between the expressions for the different boundary conditions manifest only in the sign and the $\gamma_{n}$ coefficients.

Let us reconsider the reaction probability $p_{reac}$ expressed in terms of the radial 2D Green's function to relate it with the radiation boundary constant. Using the derived short time expansions and \cite{carslaw1986conduction}
\begin{equation}
\sqrt\pi e^{x^{2}}\text{erfc}(x) = \dfrac{1}{x} - \dfrac{1}{2x^{3}} + \cdots 
\end{equation}
one obtains for sufficiently small times 
\begin{equation}
\frac{g_{\text{rad}}(R, \tau\vert R_{0})-g_{\text{abs}}(R, \tau\vert R_{0})}{g_{\text{ref}}(R, \tau\vert R_{0})-g_{\text{abs}}(R, \tau\vert R_{0})} = 1 - \tilde{h}\left(\frac{2\tau}{R+R_{0} -2} +\cdots\right)
\end{equation} 
and hence, the reaction probability defined in section \ref{reacProb} can be expressed to first order in the time step as
\begin{equation}
p_{\text{reac}} = \tilde{h}\frac{2\tau}{R+R_{0}-2} + \ldots = \kappa_{a}\frac{t}{\pi a(r+r_{0}-2a)} +\ldots.
\end{equation}
Thus, for sufficiently small times the reaction probability is the ratio of the radiation boundary reaction constant $\kappa_{a}$ to the ratio of the area $a(r+r_{0}-2a)$ and the time step.

Finally, we would like to make three comments. 
First, for large $h$, the quotient $\frac{hI_{0}(qa)-qI^{\prime}_{0}(qa)}{hK_{0}(qa)-qK^{\prime}_{0}(qa)}$ is not accurately represented by its asymptotic expansion. In these cases it is advantageous to start from
\begin{equation}\label{large_h}
\frac{1-\frac{q}{h}\frac{I^{\prime}_{0}(qa)} {I_{0}(qa)} }{1-\frac{q}{h}\frac{K^{\prime}_{0}(qa)}{K_{0}(qa)}}.
\end{equation}
We can expand \eqref{large_h} as a series in $h^{-1}$
\begin{gather}
\tilde{g}_{\text{rad}} \sim \frac{1}{2\pi D} K_{0}(qr_{0})K_{0}(qr)   \frac{I_{0}(qa)}{K_{0}(qa)}  \lbrace 1 - \frac{q}{h} \left[\frac{I_{1}(qa)}{I_{0}(qa)} + \frac{K_{1}(qa)}{K_{0}(qa)} \right] +\\
\left(\frac{q}{h}\right)^{2}  \frac{K_{1}(qa)}{K_{0}(qa)}\left[\frac{I_{1}(qa)}{I_{0}(qa)} + \frac{K_{1}(qa)}{K_{0}(qa)} \right]  - \left(\frac{q}{h}\right)^{3}  \left(\frac{K_{1}(qa)}{K_{0}(qa)}\right)^{2}\left[\frac{I_{1}(qa)}{I_{0}(qa)} + \frac{K_{1}(qa)}{K_{0}(qa)}\right] +\ldots\rbrace.
\end{gather}
Still, there can be ranges of $h$ where both expansions fail. In general, however, these ranges are small and do not pose a serious obstacle to the described method.

Second, although we only considered the $g_{\text{bc}}$ parts, small time expansions for the complete Green's functions can easily be found by using the known expressions for the free-space Green's functions, cp \eqref{laplaceAnsatz} and \eqref{laplaceFree}.  

Third, in principle, the small time expansion technique described here for the integrals involving the function $C_{n=0}$ \eqref{fRad} can also be applied for the integrals with $C_{n \neq 0}$. However, for larger $n$ the convergence behaviors of asymptotic expansions worsen, especially for $r, r_{0}$ close to $a$. Still, these expansions can prove useful for very small time steps and $r, r_{0} \gg a$.  

\section{Appendix: Relation to first-passage time simulation algorithms}\label{appendix}
As mentioned in the introduction, the presented algorithm may be regarded as a coarse-grained version of an event-driven first-passage time simulation algorithm. Here, we are demonstrating this relationship explicitly. The line of reasoning in the sections \ref{encProb}, \ref{prop} and \ref{reacProb} emphasizes the crucial role of the first-passage time. Neglecting events $\lbrace W_{t} \leq x, \tau_{a} \leq t \rbrace$ leads to the $\sqrt{\Delta t}$ error in the naive BD simulation. The conditional probability $p_{\text{enc}}$ can be used to detect these events, so that the remaining error in the formalism given above is due to the uncertainty with regard to the first-passage time. The detection of an encounter event via $p_\text{enc}$ solely provides an upper bound, i.e. $\Delta \tau_{a} = \Delta t $. On the other hand, the exact knowledge of the first-passage time would allow to set up an event-driven algorithm, with time steps equal to sampled first-passage time. It turns out that one can determine the first-passage time with any desired accuracy and without explicitly using first-passage time distributions, but only with the expressions derived so far. A similar construction has been described in \cite{ottinger1996stochastic} to determine the last reflection time. 

Consider a 1D Brownian motion, describing a molecule in the vicinity of an boundary with $W_{t_{0}} = 0$. After one simulation time step it is found at $x_{f}=W_{t_{0}+\Delta t}$. As we have just seen in \eqref{bayes1D}, $p_{\text{enc}}$ can be employed to decide if there was at least one encounter. Assuming now that there was at least one encounter, it follows $t_{0} < \tau_{a} < t_{0}+\Delta t$. The idea is now by combining iterative bisection of the time interval $[t_{0}, t_{0}+\Delta t]$ and application of $p_{\text{enc}}$ to approximate the first-passage time $\tau_{a}$ with in principle arbitrary precision. More precisely, one considers the time intervals $[t_{0}, t_{0}+ \Delta t/2]$ and $[t_{0}+\Delta t/2, t_{0} + \Delta t]$. Application of \eqref{bayes1D} requires the reconstruction of the intermediate point $x_{m} = W_{t_{0}+\Delta t/2}$  the molecule assumed at $t = t_{0} + \Delta t/2$, given that initially it was at $W_{t_{0}} = x_{i}$ and ended up at $W_{t_{0}+\Delta t} = x_{f}$. Obviously, it is not correct to sample the intermediate position according to the free-space Green's function. Instead, one considers the increments $\xi_{i} = W_{t_{0}+\Delta t /2} - W_{t_{0}}$, $\xi_{m} = W_{t_{0}+\Delta t } - W_{t_{0}+\Delta t /2}$ and $\xi_{f} = \xi_{i}+\xi_{m}$ which are by definition Gaussian random variables with $\langle \xi_{i} \rangle = \langle \xi_{m} \rangle = 0$ and $\langle \xi_{i}^{2} \rangle = \langle \xi^{2}_{m} \rangle = \sigma^{2}:=\Delta t / 2$, hence $ \langle \xi_{f} \rangle = 0$ and $\langle \xi_{f}^{2} \rangle = 2\sigma^{2} = \Delta t$.
Thus, the conditional probability density is again a Gaussian where the first moment equals half the distance between initial and terminal point  
\begin{equation}\label{midPGauss}
p(\xi_{i} = x_{m} \vert \xi_{f} = x_{f}) = \frac{p(\xi_{i} = x_{m}, \xi_{f} = x_{f})}{p(\xi_{f} = x_{f}) } = \frac{1}{\sqrt{\pi\sigma}}\exp(-\frac{(x_{m} - x_{f}/2)^{2}}{\sigma^{2}})
\end{equation}        
By sampling according to \eqref{midPGauss} (or the higher dimensional equivalent) one can construct the intermediate point. Once the intermediate point is known, one can use \eqref{bayes1D} to decide if there was an encounter in $[t_{0}, t_{0}+\Delta t  / 2]$. If there was an encounter, it follows that $t_{0} < \tau_{a} < t_{0}+\Delta t  / 2$ and we repeat the described procedure for the intervals $[t_{0}, \Delta t/2]$ and $[x_{i}, x_{m}]$. If there was no encounter, it follows that $t_{0}+\Delta t/2 < \tau_{a} < t_{0} + \Delta t $ and we repeat the procedure for the intervals $[\Delta t/2, \Delta t]$ and $[x_{m}, x_{f}]$. 

Thus, the chosen time step of the simulation corresponds to a cut-off of the first-passage time uncertainty and, conceptually, one might consider a family of the presented algorithm $\lbrace\text{SimAlg}\rbrace_{\Delta t}$, "indexed" by different time steps as a sequence of renormalization group transformations. In this picture, the limiting case with vanishing uncertainty corresponds to an exact event-driven first-passage time formalism. We would like to point out that these considerations might not only be of pure conceptual interest. It is conceivable that the given method could be used to resolve certain time periods of the simulation with finer detail. Moreover, using the algorithm as part of a hybrid algorithm, simulations on finer and coarser scales could be done in a more controlled way.   

\subsection*{Acknowledgments}
This research was supported by the Intramural Research Program of the NIH, National Institute of Allergy and Infectious Diseases. 

We would like to thank Bastian R. Angermann and Frederick Klauschen for helpful and stimulating discussions.

\bibliographystyle{plain} 
\bibliography{GreenBD}

\begin{thebibliography}{10}

\bibitem{abramowitz1964handbook}
M.~Abramowitz and I.A. Stegun.
\newblock {\em Handbook of Mathematical Functions with Formulas, Graphs, and
  Mathematical Tables}.
\newblock Dover, New York, 1965.

\bibitem{Agmon:1984p31}
N.~Agmon.
\newblock {\em J. Chem. Phys.}, 81:2811, 1984.

\bibitem{Agmon:1990p10}
N.~Agmon and A.~Szabo.
\newblock {\em J. Chem. Phys.}, 92:5270, 1990.

\bibitem{ander2004smartcell}
M.~Ander, P.~Beltrao, B.~Di~Ventura, J.~Ferkinghoff-Borg, M.~Foglierini,
  A.~Kaplan, C.~Lemerle, I.~Tomas-Oliveira, and L.~Serrano.
\newblock {\em Syst. Biol}, 1:129, 2004.

\bibitem{Andrews:2004p84}
S.S. Andrews and D.~Bray.
\newblock {\em Phys. Biol.}, 1:137, 2004.

\bibitem{BarSchieber:2002BS}
T.M.A.O.M. Barenbrug, E.A.J.F.~(Frank) Peters, and J.D. Schieber.
\newblock {\em J. Chem. Phys.}, 117:9202, 2002.

\bibitem{bender1978advanced}
C.M. Bender and S.A. Orszag.
\newblock {\em Advanced Mathematical Methods for Scientists and Engineers}.
\newblock McGraw-Hill, New York, 1978.

\bibitem{carslaw1986conduction}
H.S. Carslaw and J.C. Jaeger.
\newblock {\em Conduction of Heat in Solids}.
\newblock Clarendon Press, New York, 1986.

\bibitem{collins1949diffusion}
F.C. Collins and G.E. Kimball.
\newblock {\em J. Colloid Sci.}, 4:425, 1949.

\bibitem{deuflhard2003numerical}
P.~Deuflhard and A.~Hohmann.
\newblock {\em Numerical Analysis in Modern Scientific Computing: An
  Introduction}.
\newblock Springer, New York, 2003.

\bibitem{Edelstein:1993}
A.L. Edelstein and N.~Agmon.
\newblock {\em J. Chem. Phys.}, 99:5396, 1993.

\bibitem{einstein1956investigations}
A.~Einstein.
\newblock {\em Investigations on the Theory of the Brownian Movement}.
\newblock Dover. New York, 1956.

\bibitem{Ermak:1978p40}
D.L. Ermak and J.A. McCammon.
\newblock {\em J. Chem. Phys.}, 69:1352, 1978.

\bibitem{Gabdoulline:2002p49}
R.R. Gabdoulline and R.C. Wade.
\newblock {\em Curr. Opin. Struct. Biol.}, 12:204, 2002.

\bibitem{gillespie1976general}
D.T. Gillespie.
\newblock {\em J. Comput. Phys.}, 22:403, 1976.

\bibitem{gillespie1977exact}
D.T. Gillespie.
\newblock {\em J. Phys. Chem.}, 81:2340, 1977.

\bibitem{gillespie2001approximate}
D.T. Gillespie.
\newblock {\em J. Chem. Phys.}, 115:1716, 2001.

\bibitem{green1988simulation}
N.J.B. Green.
\newblock {\em Mol. Phys.}, 65:1399, 1988.

\bibitem{harrison1985brownian}
J.M. Harrison.
\newblock {\em Brownian Motion and Stochastic Flow Systems}.
\newblock Wiley, New York, 1985.

\bibitem{Hattne:2005p86}
J.~Hattne, D.~Fange, and J.~Elf.
\newblock {\em Bioinformatics}, 21:2923, 2005.

\bibitem{kampen2007stochastic}
N.G. Kampen.
\newblock {\em Stochastic Processes in Physics and Chemistry}.
\newblock Elsevier, New York, 2007.

\bibitem{Lamm:1984p4}
G.~Lamm.
\newblock {\em J. Chem. Phys.}, 80:2845, 1984.

\bibitem{Lamm:1983p12}
G.~Lamm and K.~Schulten.
\newblock {\em J. Chem. Phys.}, 78:2713, 1983.

\bibitem{morelli2008reaction}
M.J. Morelli and P.R. ten Wolde.
\newblock {\em J. Chem. Phys.}, 129:054112, 2008.

\bibitem{North:1984}
S.H. Northrup, S.A. Allison, and A.~McCammon.
\newblock {\em J. Chem. Phys.}, 80:1517, 1984.

\bibitem{NorthCurv:1986}
S.H. Northrup, M.S. Curvin, S.A. Allison, and A.~McCammon.
\newblock {\em J. Chem. Phys.}, 84:2196, 1986.

\bibitem{Oppelstrup:2009p144}
T.~Oppelstrup, V.V. Bulatov, A.~Donev, M.H. Kalos, G.H. Gilmer, and B.~Sadigh.
\newblock {\em Phys. Rev. E}, 80:066701, 2009.

\bibitem{Opplestrup:2006p235}
T.~Opplestrup, V.V. Bulatov, G.H. Gilmer, M.H. Kalos, and B.~Sadigh.
\newblock {\em Phys. Rev. Lett.}, 97:230602, 2006.

\bibitem{ottinger1996stochastic}
H.C. {\"O}ttinger.
\newblock {\em Stochastic Processes in Polymeric Fluids: Tools and Examples for
  Developing Simulation Algorithms}.
\newblock Springer, Berlin, 1996.

\bibitem{pimblott1992stochastic}
S.M. Pimblott and N.J.B. Green.
\newblock {\em J. Phys. Chem.}, 96:9338, 1992.

\bibitem{redner2001guide}
S.~Redner.
\newblock {\em A Guide to First-passage Processes}.
\newblock Cambridge University Press, Cambridge, 2001.

\bibitem{Rice:1985}
S.~A. Rice.
\newblock {\em Diffusion Limited Reactions}.
\newblock Elsevier, New York, 1985.

\bibitem{risken1996fokker}
H.~Risken.
\newblock {\em The Fokker-Planck Equation: Methods of Solution and
  Applications}.
\newblock Springer, New York, 1996.

\bibitem{rodriguez2006spatial}
J.V. Rodr{\'\i}guez, J.A. Kaandorp, M.~Dobrzy{\'n}ski, and J.G. Blom.
\newblock {\em Bioinformatics}, 22:1895, 2006.

\bibitem{shoup1982role}
D.~Shoup and A.~Szabo.
\newblock {\em Biophys. J.}, 40:33, 1982.

\bibitem{schutter2001computational}
J.R. Stiles.
\newblock {\em Computational Neuroscience: Realistic Modeling for
  Experimentalists}.
\newblock CRC Press, Boca Raton, 2001.

\bibitem{Takahashi:2010p139}
K.~Takahashi, S.~T{\u a}nase-Nicola, and P.R. ten Wolde.
\newblock {\em Proc. Natl. Acad. Sci. U.S.A.}, 107:2473, 2010.

\bibitem{vanZon:2005p340}
J.S. van Zon and P.R. ten Wolde.
\newblock {\em Phys. Rev. Lett.}, 94:128103, 2005.

\bibitem{vanZon:2005p401}
J.S. van Zon and P.R. ten Wolde.
\newblock {\em J. Chem. Phys.}, 123:234910, 2005.

\end{thebibliography}

\end{document}